\documentclass[letterpaper]{article} 
\usepackage{aaai24}   
\usepackage{times}  
\usepackage{helvet}  
\usepackage{courier}  
\usepackage[hyphens]{url}  
\usepackage{graphicx} 
\usepackage{natbib}  
\usepackage{caption} 
\usepackage{algorithm}
\usepackage{algorithmic}
\usepackage{multirow}

\usepackage{amsmath}

\usepackage{xcolor}

\usepackage{newfloat}
\usepackage{listings}
\DeclareCaptionStyle{ruled}{labelfont=normalfont,labelsep=colon,strut=off} 
\floatstyle{ruled}
\newfloat{listing}{tb}{lst}{}
\floatname{listing}{Listing}
\title{Unpacking Discourses on Childbirth and Parenthood in Popular Social Media Platforms Across China, Japan, and South Korea}

\title{Unpacking Discourses on Childbirth and Parenthood in Popular Social Media Platforms Across China, Japan, and South Korea}

\author {
    Zheng Wei\textsuperscript{\rm 1},
    Yunqi Li\textsuperscript{\rm 2},
    Yucheng He\textsuperscript{\rm 2},
    Yuelu Li\textsuperscript{\rm 2},
    Xian Xu\textsuperscript{\rm 3},
    Huamin Qu\textsuperscript{\rm 1},
    Pan Hui\textsuperscript{\rm 2},
    Muzhi Zhou\textsuperscript{\rm 2}
}
\affiliations {
    \textsuperscript{\rm 1}The Hong Kong University of Science and Technology\\
    \textsuperscript{\rm 2}The Hong Kong University of Science and Technology (Guangzhou)\\
    \textsuperscript{\rm 3}Lingnan University\\
    zwei302@connect.ust.hk, yli895@connect.hkust-gz.edu.cn, yhe772@connect.hkust-gz.edu.cn,  yli883@connect.hkust-gz.edu.cn, xianxu@ln.edu.hk, huamin@cse.ust.hk, panhui@hkust-gz.edu.cn, mzzhou@hkust-gz.edu.cn
}

\usepackage{bibentry}

\begin{document}

\maketitle

\begin{abstract}
Social media use has been shown to be associated with low fertility desires. However, we know little about the discourses surrounding childbirth and parenthood that people consume online. We analyze 219,127 comments on 668 short videos related to reproduction and parenthood from \textit{Douyin} and \textit{Tiktok} in China, South Korea, and Japan, a region famous for its extremely low fertility level, to examine the topics and sentiment expressed online. \textit{BERTopic} model is used to assist thematic analysis, and a large language model \textit{QWen} is applied to label sentiment. We find that comments focus on \textit{childrearing costs} in all countries, \textit{utility of children}, particularly in Japan and South Korea, and \textit{individualism}, primarily in China. Comments from \textit{Douyin} exhibit the strongest anti-natalist sentiments, while the Japanese and Korean comments are more neutral. Short video characteristics, such as their stances or account type, significantly influence the responses, alongside regional socioeconomic indicators, including GDP, urbanization, and population sex ratio. This work provides one of the first comprehensive analyses of online discourses on family formation via popular algorithm-fed video sharing platforms in regions experiencing low fertility rates, making a valuable contribution to our understanding of the spread of family values online.
\end{abstract}



\section{Introduction}

{The widespread use of social media among young people, along with its capacity to disseminate gender and family values, has significant implications for the future population structure. \textit{TikTok}, for example, being ranked as the 5th most popular social app globally in 2024, has 70.10\% of its users between the ages of 18 and 34~\cite{TK01}.} Online discourses can affect people's everyday practices and has a significant influence on people's value formation~\cite{papacharissi2015affective, perloff2014social,wei2024digital}. As global population trends are increasingly influenced by low fertility rates, scholars are beginning to explore the connection between fertility intentions and online activities. Many studies have indicated that social media use is linked to reduced fertility desire or fertility levels~\cite{liu2021influence,wildeman2023fertility}. 

Despite the significant potential of social media in shaping norms and cultural expectations about marriage and family, there are few analyses of the popular online discourse surrounding childbirth and parenthood. We know little about the fertility and family-related discourses that are disseminated on popular social media platforms. This lack of understanding about these online discourses significantly hinders our comprehension of the relationship between social media consumption, family ideals, and low fertility. 


Low fertility rates have persisted over decades in China, Japan, and South Korea. In 2022, the Total Fertility Rates (TFR), indicating the average number of children a woman would have during her lifetime, reached 1.23 in China, 1.26 in Japan, and 0.78 in South Korea~\cite{RN3, RN4}. 
Understanding how the public perceives childbirth and parenthood is crucial for understanding this lowest-low fertility pattern. Taking Japan as an example, the government has been promoting pro-natalist policies like financial incentives and childcare support {since the early 1990s.} However, these measures had little impact on fertility level \cite{boling1998family}. A similar lack of policy response is observed in China, {with the relaxation of the one-child policy since 2013}. The persistent low-fertility trends imply that this region appears to have adopted a low-fertility norm. This norm reflects a consensus about families with one or zero child, or, even anti-natalist views \cite{fernandez2009culture}. Anti-natalism is a philosophical position that holds a negative view on procreation \cite{smyth2020question}. In ``\textit{Better Never to Have Been},'' philosopher David Benatar argues that bringing a child into the world inevitably subjects it to significant suffering, given the unsatisfactory state of the world \cite{benatar2008better}. We suspect that this anti-natalist sentiment may dominate those discourses on popular social media.

In this paper, we examine content related to childbirth and parenthood on popular social media platforms and how this content influences online discussions. We also explore how short videos may provoke different responses. The interactions between content creators and viewers, along with the socioeconomic context and online expressions, reveal how social media platforms influence the information consumed by more and more users. This is a key theme of ICWSM. The specific research questions are:

RQ1: What do people discuss when watching childbirth or parenthood-related short videos? 

{RQ2: What short video characteristics influence the discussion sentiments about childbirth and parenthood?}

RQ3: What regional-specific socioeconomic factors shape those discussions about childbirth and parenthood?

We analyzed over 200k online comments from 668 short videos about childbirth and parenthood in China, Japan, and South Korea from popular platforms like \textit{Douyin} and \textit{TikTok}. We used \textit{BERTopic} algorithm to assist topic analysis and Large Language Model (LLM) \textit{QWen} to label sentiments toward fertility. To understand how these discourses are shaped by various videos, we group the videos based on the account types and whether the content is pro- or against reproduction. To understand how these discourses are shaped by local socioeconomic context, we link the comments' geolocation to its local socioeconomic indicators, such as local GDP, at a provincial (for China) or national level (for Japan and South Korea) \cite{zhou2024can,wei2024social}. 

This study provides one of the first insights into how childbirth and parenthood are interpreted and perceived on popular social media platforms. This research contributes to understanding the impact of the Web in shaping perceptions and narratives around childbirth and parenthood in a region known for its lowest fertility rates globally. This study offers a new perspective to understand the global demographic shift to low fertility and population aging.

\section{BACKGROUND AND THEORIES}

\subsection{Low Fertility in East Asia}

China, Japan, and South Korea are neighboring East Asian nations that have experienced rapid industrialization and modernization. This region is also deeply influenced by Confucian values, 
emphasizing social order and family hierarchy. Typical values include men's superiority, respect for authority, and a commitment to education \cite{sin2012confucianism}. Most urban families in China have only one child, while Japan and South Korea have some of the highest rates of childlessness in the world. Approximately three in ten women in Japan and one in four women in South Korea are permanently childless \cite{frejka2010east}.

\subsubsection{China}

The introduction of the one-child policy in 1979 was followed by a sharp reduction in birth rates \cite{settles2012one}. Despite the policy's relaxation to allow two children in 2016 and further to three children in 2021, fertility rates have continued to drop \cite{zhang2022fertility}. Elevated education levels and career aspirations, especially of women, are accompanied by delayed marriage, especially in urban areas with rising housing prices \cite{zhang2023women}.

\subsubsection{Japan}
 Economic instability and high living costs have been regarded as one of the leading reasons behind the extremely low fertility \cite{allison2014precarious}. This situation is exacerbated by the overtime working culture \cite{chandra2012work}. Traditional gender roles also persist, with the majority of women leaving the labor market once married or having children \cite{brinton2023women}.

\subsubsection{South Korea}

South Korea's fertility rate has plummeted to one of the lowest in the world, driven by factors similar to those in China and Japan \cite{lee2015lowest}. High costs associated with education, housing, along with intense academic competition, place substantial pressure on parents \cite{chin2014family}.

\subsection{Interpreting childbirth and parenthood}
We summarize key literature to form a five-dimensional framework (health, population structure, child-rearing cost, utility of family, and values) to explain family formation in East Asia to guide our later analysis.

\subsubsection{Childrearing Cost}
Parents are expected to provide good education, healthcare, and housing to nurture their children to become ``useful'' \cite{zhang2020tiger}. Exorbitant living expenses and intense educational competition exacerbate the difficulty of meeting this expectation \cite{fisher2019balancing}. Subjective economic pressures, such as feelings of financial scarcity and uncertainty, stem from a pessimistic view of the future \cite{chamon2013income}. This sense of the undesirable future lays the foundation for anti-natalist views. 


\subsubsection{Utility}
Having children in Confucianism can fulfill three main responsibilities: supporting one's parents, showing them respect, and continuing the family lineage \cite{tang1995confucianism}. Among these, continuing the family lineage is the most important. Moreover, elderly care still largely depends on adult children \cite{zhou2022intergenerational}. Beyond Confucianism family values, having children fosters a sense of intimacy and provides emotional rewards, contributing to the well-being of parents \cite{nomaguchi2020parenthood}. Therefore, the important functions that families play provide strong support for a pro-natalist view.

\subsubsection{Values}
Confucianism asserts that ``There are three unfilial acts, and having no descendants is the greatest,'' underscoring the deep-rooted conviction that having children is essential in a collective context \cite{tang1995confucianism}. The saying "More children, more blessings" reflects a traditional belief that having more children is a positive value. Those who defend these views may be seen as upholding outdated ideas.

Rapid industrialization, urbanization, and economic growth have introduced modern values such as individualism, self-realization, and personal autonomy. Young adults increasingly prioritize personal goals, including higher education, career advancement, and personal fulfillment over marriage and childbearing \cite{lesthaeghe2011second}. 

Family values in East Asia appear to blend individualistic and traditional beliefs, highlighting the importance of both personal development and family continuity \cite{inglehart2000modernization}. Meanwhile, gender roles at home remain traditional. Although women have made significant advances in education and workforce participation, they are expected to be the prime caregivers \cite{kan2022revisiting}. The burden of balancing professional aspirations with domestic responsibilities discourages many women from marrying or having children \cite{ ochiai2011unsustainable}. 


\subsubsection{Population Structure}

An aging population is linked to labor shortages, which are key motivations for pro-natalist policies \cite{demeny1986pronatalist}. A rapidly aging population places significant pressure on old-age spending. Low fertility has weakened intergenerational support structures. In East Asian societies that emphasize filial piety, the expectation to be cared for by adult children can impose a high level of anxiety due to low fertility \cite{lin2013comparative}. Discussions about concerns related to a shrinking population, labor shortage, and elderly care reflect a pro-natal sentiment, while support for greater independence in self-care or increased public assistance for elderly care indicates a supportive attitude towards low fertility.

\subsubsection{Health}

Health and medical care are also related to childbirth. Anxiety and fears of childbirth are common among childbearing women. Concerns about labor pain and complications can lead to remaining childless, bolstering anti-reproduction sentiment \cite{dencker2019causes}. In contrast, access to assisted reproductive technology (ART), such as in-vitro fertilization (IVF), enables infertile people to have children \cite{yao2023ambivalence}. Discussions related to ART should demonstrate a pro-natal sentiment.

\subsection{Short Videos and Responses}

Short video content has a huge impact on related discussions. The \textit{S-O-R (Stimulus-Organism-Response)} framework is a classic psychological model to analyze how stimuli from the environment influence an individual's internal processes (organism) and lead to a specific response \cite{mehrabian1974approach}. In social media analysis, this model helps understand how various types of content (stimulus) affect users' thoughts, feelings, and behaviors (organism), ultimately resulting in their reactions (response) \cite{xu2021reaching}.

Content creators' identity is the prime stimulus. \textit{social relation theory} suggests that the social distance between communicators and audiences affects the reception of information \cite{haridakis2009social}. Personal accounts, perceived as peer communicators with small social distance, enhance emotional resonance and information acceptance among audiences. In contrast, mainstream or official narratives often use grand rhetoric to promote reproductive policies. This increases the social distance between content creators and their audiences, which can lead to backlash and result in less positive support for pro-natalist policies.

\subsection{Socioeconomic Context and Responses}
Viewers respond to the same information differently based on their own experiences and considerations of childbearing. The cost of childbirth is a significant factor contributing to low fertility. \textbf{economic development}, usually represented by GDP, can lead to increased purchasing power, making the high costs associated with childbirth more manageable. \textbf{Urbanization} is linked to rising housing prices \cite{jeanty2010estimation}, which strongly correlate with declining marriage rates. 

Values play a crucial role. \textit{Modernization theory} and the \textit{Second Demographic Transition Theory} suggest that the emergence of more individualistic values results in lower fertility rates and more diverse family structures \cite{kolpashnikova2020country}. 
Research in Western societies indicates that fertility rates are generally higher in areas with more gender-equal social norms \cite{goldscheider2015gender}. It is uncertain whether this trend applies to East Asia. A distorted \textbf{male-to-female sex ratio} is a clear sign of sex selection at birth and reflects a low level of gender equality \cite{das2003preference} and may lead to modern women's strong attitudes against childbearing.

\section{DATA}
Data are from \textit{Douyin} for Chinese comments and \textit{TikTok} for Japanese and South Korean comments. Both are algorithm-led short-form video sharing platforms. \textit{Douyin}, being China's leading short video platform, has over 1 billion registered users and more than 700 million daily active users~\cite{TK01}. \textit{TikTok}, the global version of \textit{Douyin}, was the most‑downloaded app in 2024 and had 1.6 billion active users \cite{TK01}. {The average age of \textit{Douyin} users is 28 years (60\% female; 78\% aged 18–35). \textit{TikTok} Japan users have a mean age of 27 years, and 52\% are females. \textit{TikTok} South Korea is gender-balanced, with the users primarily of 18–24-year-olds and an average age of 25 years \cite{TK01}.} 

Short videos usually have multiple hashtags to reflect their relevant topics for promotional reasons. We located related short videos through hashtag keyword search. {Specifically, we adopted a hashtag-snowball strategy to retrieve short videos, similar to one previous study \cite{wei2024social}. First, an expert in global family studies selected 10 seed keywords in English (1. single, 2. low fertility, 3. childrearing pressure, 4. childless, 5. DINK, 6. young parents, 7. childbirths, 8. family prosperity, 9. big family, 10. family policy). We translated these keywords into the corresponding language and input them into the platform, and the platform would offer suggested terms reflecting trending hagtags related to the seed keywords. For example, if ``single'' or ``childbirth'' were input in the search box, the platform's search system would automatically recommend tags like ``no marriage'' and ``natural labor,'' respectively. We added the most recommended tags to each seed keyword to reach a list of 20 high-frequency hashtag keywords per country (in Appendix Table \ref{A1}). Including more hashtag words leads to inefficient searches because a single short video often has multiple hashtags. This means that adding more highly correlated hashtags can cause the same video to be crawled repeatedly. Ultimately, we crawled short videos tagged with any of these keywords and their corresponding comments. Using the same seed keywords and then modifying them based on the platform-specific algorithms ensures comparable yet platform-specific videos across China, Japan, and South Korea to reflect the information consumed by users.}

To collect corresponding comments, {we modified the web crawler from the open-source tool \textit{MediaCrawler}\footnote{\url{https://github.com/NanmiCoder/MediaCrawler}}. We stopped comment collection once each country reached approximately 100,000 comments (10 hrs/day, 15 days for China and 19 days for Japan and Korea). This number is similar to but larger than the sample size used in the most recent social media content analysis studies \cite{plepi2024perspective,aleksandric2024users}. We also noted that further crawling resulted in a growing proportion of duplicated comments and became less efficient.} 

Due to China's significantly larger user base compared to those in Japan and Korea, there is a much higher volume of comments per video in China. Therefore, the number of videos in China is relatively small. {We compared the video characteristics of Japan and South Korea, which were crawled over the same 15 days as those for China, along with the full 19 days of data (reported in Table \ref{distribution} vs. Appendix Table \ref{Appendix: reducedvideosample}). Video characteristics from the two crawling durations are highly similar.}

The crawled videos span from {April 2019 to August 2024} across all three countries. For China, we removed comments posted before April 2022 because \textit{Douyin} began attaching geotags to all content since April 2022 \cite{10.1145/3704991.3704994}. We removed comments containing only punctuation marks, emojis, or @usernames, leaving 288,101 comments. {Table \ref{appendix:A2-XXX} presents the step-by-step selection of videos and comments. In the end, we analyzed 219,127 comments from 668 videos (China = 74,828 (Apr 2022–Aug 2024); Japan = 82,887 (Apr 2019–Aug 2024); South Korea = 61,412 (Apr 2019–Aug 2024).}

\begin{table}[htp]
\centering
\caption{Video and comment numbers by data-filtering stage}
\scalebox{0.68}{
\begin{tabular}{cccccc}
\hline
Stages                                                                                                                                & Data type                                                 & CN                                                    & JP                                                     & KR                                                     & Total                                                  \\ \hline
\begin{tabular}[c]{@{}c@{}}I. Keyword \\ Search\end{tabular}                                                                          & \begin{tabular}[c]{@{}c@{}}Videos\\ Comments\end{tabular} & \begin{tabular}[c]{@{}c@{}}126\\ 137,861\end{tabular} & \begin{tabular}[c]{@{}c@{}}2077\\ 226,073\end{tabular} & \begin{tabular}[c]{@{}c@{}}2648\\ 233,171\end{tabular} & \begin{tabular}[c]{@{}c@{}}4851\\ 597,105\end{tabular} \\ \hline
\begin{tabular}[c]{@{}c@{}}II. Removed duplicated \\ and irrelevant 
 videos\end{tabular}                                                & \begin{tabular}[c]{@{}c@{}}Videos\\ Comments\end{tabular} & \begin{tabular}[c]{@{}c@{}}114\\ 136,125\end{tabular} & \begin{tabular}[c]{@{}c@{}}317\\ 114,826\end{tabular}  & \begin{tabular}[c]{@{}c@{}}246\\ 104,996\end{tabular}  & \begin{tabular}[c]{@{}c@{}}677\\ 355,947\end{tabular}  \\ \hline
\begin{tabular}[c]{@{}c@{}}III. Removed duplicates \\ and punctuation/emoticon/\\ @mention-only comments\end{tabular} & \begin{tabular}[c]{@{}c@{}}Videos\\ Comments\end{tabular} & \begin{tabular}[c]{@{}c@{}}113\\ 95,533\end{tabular}  & \begin{tabular}[c]{@{}c@{}}317\\ 112,777\end{tabular}  & \begin{tabular}[c]{@{}c@{}}246\\ 79,791\end{tabular}   & \begin{tabular}[c]{@{}c@{}}676\\ 288,101\end{tabular}  \\ \hline
\begin{tabular}[c]{@{}c@{}}IV. Removed off-topic \\word clusters after BERT\end{tabular}                                                            & \begin{tabular}[c]{@{}c@{}}Videos\\ Comments\end{tabular} & \begin{tabular}[c]{@{}c@{}}105\\ 74,828\end{tabular}  & \begin{tabular}[c]{@{}c@{}}317\\ 82,887\end{tabular}   & \begin{tabular}[c]{@{}c@{}}246\\ 61,412\end{tabular}   & \begin{tabular}[c]{@{}c@{}}668\\ 219,127\end{tabular}  \\ \hline
\end{tabular}
}
\label{appendix:A2-XXX} 
\end{table}

\section{METHODS}
\subsection{{Short-video Classification} 
 \label{m-videocoding} }
Three researchers, who can read and speak Chinese, Japanese, and Korean, respectively, reviewed all videos and coded the following video characteristics. One senior researcher randomly selected 10\% of the videos from each country to verify the short video coding. 

First, short video account types include (i) parents sharing personal experiences, (ii) bystanders commenting on others' childbirth or parenthood experiences, (iii) experts (government officers or family scholars), and (iv) mainstream media. The stance toward childbirth of the specific video is further coded as pro-reproduction or anti-reproduction. 
 
The main figure displayed in the short video is coded into their gender and age group (young (approximately 18-30 years old), middle-aged (approximately 31-55 years old), old (approximately 56+ years old)); if the characteristics of the main figure cannot be identified, e.g., news reporting of a family policy, we use the voice-over to infer gender and age. This information is useful to answer RQ2, where we would like to understand how short video characteristics drive discussions.

\begin{table}[ht]
\centering
\caption{Distribution of short video characteristics.}
\scalebox{0.75}{
\begin{tabular}{ccccccc}
\hline
Video Characteristics                                                                & CN  & Total                & JP  & Total                & \begin{tabular}[c]{@{}c@{}}SK \end{tabular} & Total                \\ \hline
\begin{tabular}[c]{@{}c@{}} Anti-reproduction\end{tabular} & 61\%   & \multirow{2}{*}{105} & 51.1\% & \multirow{2}{*}{317} & 41.9\%                                                & \multirow{2}{*}{246} \\
\begin{tabular}[c]{@{}c@{}}Pro-reproduction\end{tabular}   & 39\%   &                      & 48.9\% &                      & 58.1\%                                                &                      \\ \hline
Gender-M                                                            & 36.2\% & \multirow{2}{*}{105} & 34.7\% & \multirow{2}{*}{317} & 38.6\%                                                & \multirow{2}{*}{246} \\
Gender-F                                                            & 63.8\% &                      & 65.3\% &                      & 61.4\%                                                &                      \\ \hline
Age-Young                                                           & 41.9\% & \multirow{3}{*}{105} & 44.8\% & \multirow{3}{*}{317} & 85.8\%                                                & \multirow{3}{*}{246} \\
Age-Middle                                                          & 40.9\% &                      & 51.1\% &                      & 13\%                                                  &                      \\
Age-Old                                                             & 17.2\% &                      & 4.1\%  &                      & 1.2\%                                                 &                      \\ \hline
Personal Sharing                                                    & 39\%   & \multirow{4}{*}{105} & 60.2\% & \multirow{4}{*}{317} & 41\%                                                  & \multirow{4}{*}{246} \\
Bystanders                                                          & 37.2\% &                      & 17.7\% &                      & 38.2\%                                                &                      \\
Experts                                                             & 16.2\% &                      & 15.8\% &                      & 16.3\%                                                &                      \\
Mainstream Media                                                      & 7.6\%  &                      & 6.3\%  &                      & 4.5\%                                                 &                      \\ \hline
\end{tabular}
}
\label{distribution}
\end{table}

{Table \ref{distribution} reports the distribution of these characteristics at the video level.} First, 61\% of the Chinese videos expressing anti-reproduction views. This proportion is much higher compared to Japan and South Korea. Videos predominantly feature female figures or are created by female creators, emphasizing women's central role in showcasing childbirth and parenthood-related information. Individuals featured in the videos are mostly young, typically younger than 56 years old. The majority of videos are shared by individual accounts sharing personal childbirth or parenting experiences, followed by those commenting on others' experiences. In contrast, videos from experts and mainstream media represent a small share.

\subsection{Topic Analysis}
 {Our research team includes members who can read and speak Chinese, Japanese, and Korean, respectively. Keywords and comments were retained and analyzed in their original languages--simplified Chinese, Japanese, and Korean--without translation for analyses.}

We used inductive and deductive approaches to extract major themes from comments, following the \textit{Grounded Theory} proposal \cite{corbin2014basics}. This proposal organizes texts into hierarchical layers. In the first step, comments are grouped into word clusters, and these clusters are further synthesized into broader topic themes and eventually into theoretical dimensions, drawing on thematic analysis methods demonstrated by Boyd-Graber et al. and Baumer et al.\cite{baumer2017comparing,boyd2017applications}. 

\begin{table}[ht]
\centering
\caption{Inductive and deductive phases for topic analysis}
\scalebox{0.9}{
\begin{tabular}{cccc}
\hline
Stages                                                                                          & CN & JP & \begin{tabular}[c]{@{}c@{}} KR \end{tabular} \\ \hline
I. BERTopic output: \\ number of Word Clusters                                                                                 & 245   & 291   & 302                                                   \\ \hline
\begin{tabular}[c]{@{}c@{}}II. Removing irrelavant \\ word clusters\end{tabular}                    & 219   & 196   & 220                                                   \\ \hline
\begin{tabular}[c]{@{}c@{}}III. Consolidate word clusters \\ into Topic Themes\end{tabular}          & 17    & 18    & 16                                                    \\ \hline
\begin{tabular}[c]{@{}c@{}}IV. Allocate topic themes \\ into Theoretical Dimensions\end{tabular} & 5     & 5     & 5                                                     \\ \hline
\end{tabular}
}
\label{tab-topicphase}
\end{table}

For the inductive step, we employed the \textit{BERTopic} algorithm~\cite{grootendorst2022bertopic}, a topic modeling technique that leverages the power of transformer-based embeddings, to classify a large dataset of unstructured user comments into word clusters. This method is particularly effective for analyzing social media \cite{egger2022topic} content and has been recommended and applied in recent studies~\cite{wei2024social,jeong2019social}. The algorithm was used to generate and visualize a topic distance matrix, where the distance between every pair of topics represents their relationships.

We optimized the \textit{BERTopic} model with \textsc{TopicTuner}\footnote{https://github.com/drob-xx/TopicTuner}, selecting the hyper-parameters that best balanced thematic granularity against coverage (\texttt{min\_cluster\_size/min\_samples}: CN = 92/9, JP = 74/22, KR = 70/7). With these settings, \textit{BERTopic} generated country-specific clusters and their top c–TF–IDF word representations. 
\textit{BERTopic} analysis resulted in 245 clusters from the Chinese comments, 302 clusters from the Korean comments, and 291 clusters from the Japanese comments. Given that the number of word clusters is too high for easy comprehension, we need to reduce the number of word clusters into sensible topic themes.

\begin{figure}[H]
    \centering
    \includegraphics[width=7cm]{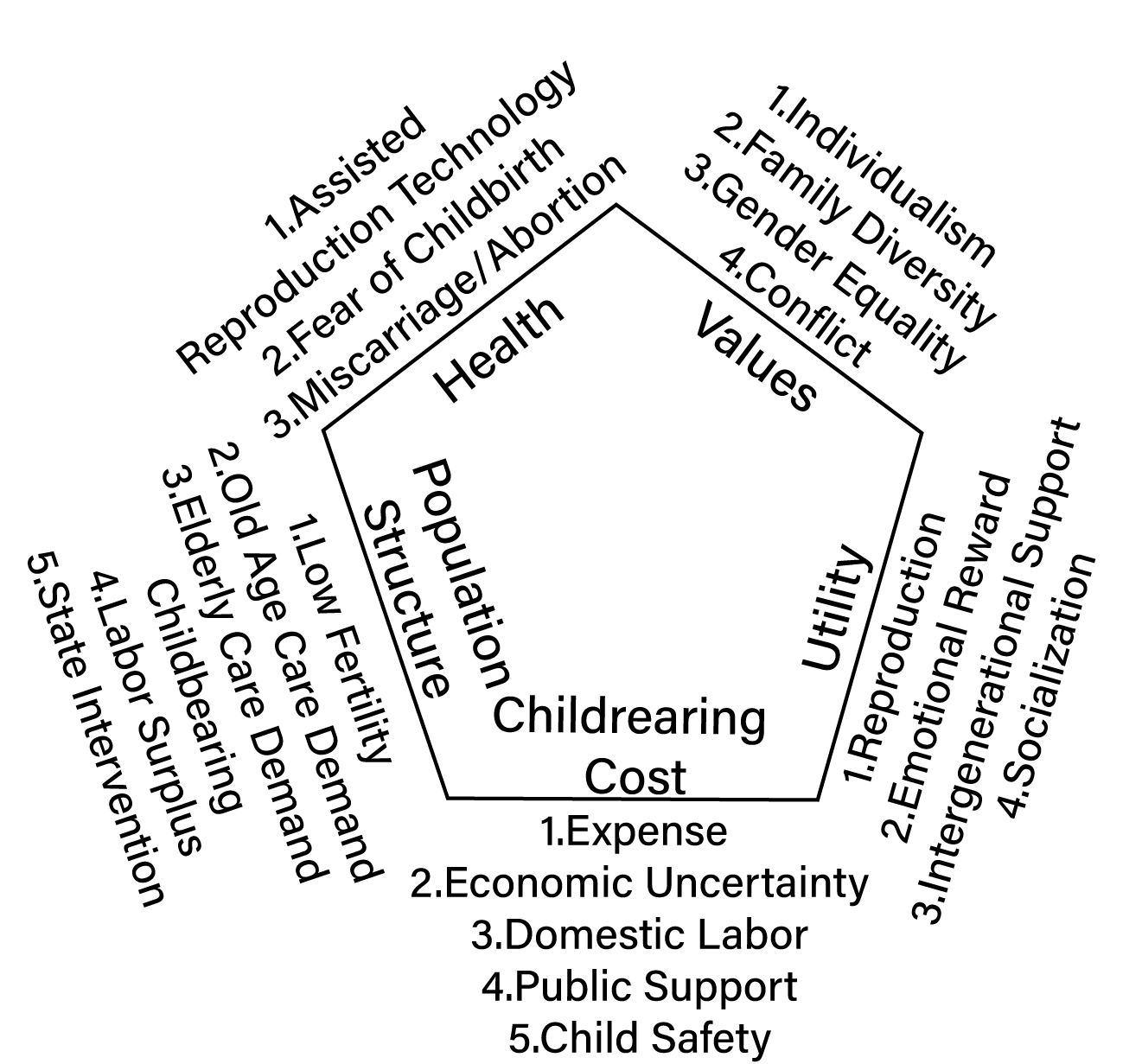}
    \caption{21 topic themes and their corresponding five theoretical dimensions.}
    \label{topic-3}
\end{figure}

In this deductive step, we merged the word clusters based on lexical synonyms,  {Stage III in Table~\ref{tab-topicphase}.} The hierarchical topic tree from the \textit{BERTopic} outputs (in the Results folder of GitHub) is used as a reference to merge the word clusters. We merged ``Fear of Childbirth'' and ``Healthcare Support'' because their branches in the topic tree figure were linked. {However, a single cut-off in this tree hierarchy still creates many fragmented topics with unrelated clusters. Therefore, we used this hierarchical tree as a reference starting point and moved backward from the top to the root of the hierarchical tree. This process is guided by the five theoretical dimensions and their related themes from the literature discussed above, with the supervision of a senior family scholar in two group meetings. Three coders then independently coded the word cluster and discussed discrepancies in further group meetings. The inter-coder agreement on average was $\kappa = 0.82$. Concise topic themes emerged from those closely related word clusters. This process follows the standard \textit{Grounded Theory} proposal \cite{corbin2014basics}.} This refinement reduced the hundreds of word clusters to 17 topic themes for China, 18 for Japan, and 16 for South Korea, as shown in Table~\ref{tab-topicphase}. Four researchers aligned these topic themes into English. The resulting 21 synthesized themes in English are presented in Figure~\ref{topic-3}.

Finally, all researchers discussed and grouped these topic themes into five theoretical dimensions--\textit{Health, Population Structure, Child-rearing Cost, Utility, and Values} (Figure~\ref{topic-3}), based on the theoretical discussions above. 

\subsubsection{Illustrative comment quotations}\label{Illustrative quotations}
{\textit{BERTopic} outputs provide word representation and the representative comment document containing the word representation for each word cluster, based on c-TF-IDF weighting. Because one topic theme contains multiple word clusters, there are multiple representative comments. When illustrating the topic themes, we selected one of the representative comment documents from the \textit{BERTopic} output as examples. We uploaded the exemplar \textit{BERTopic} output with the word representation and the representative document to GitHub.}

\subsection{LLM Assisted Sentiment Analysis}

We categorize each comment based on whether it expresses supportive, against, or neutral attitudes toward childbirth and parenthood. Many comments reflect more balanced, mixed views. For example, \textit{''Many times I think that if I had children, I would be very happy, but even without them, I am currently very happy and content.''} This nuanced classification of comments makes specified NLP models, such as the Roberta-base-cold model, less effective, and large language model (LLM) can help \cite{gilardi2023chatgpt,weicontextaware}. We used the open-sourced multilingual LLM \textit{Tongyi Qianwen} \footnote{\url{https://cn.aliyun.com/product/bailian?from_alibabacloud=}} or \textit{QWen} (in English) \footnote{\url{https://github.com/QwenLM/Qwen}}. This LLM is developed by \textit{Alibaba Cloud} and has an overall performance of its 72B model outperforms GPT-3.5 on 7 out of 10 tasks \cite{bai2023qwen}. 

\textbf{Prompt Testing} Sentiment classification was performed on the {original‐language corpora}. The prompts were iteratively refined (6 rounds) by gradually adding more examples. In each round, we evaluate the prompt performance by annotating 1,000 comments randomly selected from each country using LLM. The same 1,000 comments per language were manually coded. Using the prompt finalized in Round 6, \textit{Qwen-turbo} achieved 88–93\% alignment with human coders and macro-F1 scores within ± 0.03 for Chinese, Japanese, and Korean comments (Appendix Table \ref{sentiment-test}), {indicating comparable performance across the three corpora.}

{\textbf{Experiment Setting} Sentiment annotations are processed separately based on human-coded video stances towards reproduction (detailed in METHODS\ref{m-videocoding}). Specifically, the comment samples are divided into two groups: comments related to short videos that express an anti-reproduction stance and comments related to videos that express a pro-reproduction stance and were annotated using corresponding prompts. This setting aims to solve the issue that sentiment and stance are not always the same \cite{Bestvater_Monroe_2023}. For example, positive or supportive sentiments expressed in comments regarding videos that convey anti-reproduction views should be coded as anti-natal, while the same sentiments in response to videos that convey pro-reproduction views should be coded as pro-natal. In the following discussion, we present the `pro-natal,' 'anti-natal,' or `neutral' stances as comment sentiments for easy interpretation. }

After LLM annotated all comments, we randomly selected 1,000 comments from each country for human annotation and compared them with LLM results. Over 88\% of the comments categorized by humans match the LLM categories. Appendix Table \ref{sentiment-test2} reports the test results. {Again, the performance was comparable across languages.}

Table \ref{sentiment} presents the sentiment analysis outcome. Comments from China reveal the highest level of anti-natal sentiment, and those from Japan show the highest level of neutral sentiment. When discussing the sentiment analysis results, we dropped the less than 1\% unidentifiable sentiments.

\begin{table}[ht]
\centering
\caption{Distribution of sentiment categories by country}
\scalebox{0.85}{
\begin{tabular}{cccccc}
\hline
                                                      & Anti-natal                                                    & Pro-natal                                                    & Neutral                                                     & Unidentifiable                                         & Total  \\ \hline
CN                                                 & \begin{tabular}[c]{@{}c@{}}37,504\\ (50.12\%)\end{tabular} & \begin{tabular}[c]{@{}c@{}}14,952\\ (19.98\%)\end{tabular} & \begin{tabular}[c]{@{}c@{}}22,105\\ (29.54\%)\end{tabular}  & \begin{tabular}[c]{@{}c@{}}267\\ (0.36\%)\end{tabular} & 74,828  \\
JP                                                 & \begin{tabular}[c]{@{}c@{}}24,892\\ (30.03\%)\end{tabular} & \begin{tabular}[c]{@{}c@{}}23,757\\ (28.66\%)\end{tabular} & \begin{tabular}[c]{@{}c@{}}34,021\\ (41.05\%)\end{tabular} & \begin{tabular}[c]{@{}c@{}}217\\ (0.26\%)\end{tabular} & 82,887 \\
\begin{tabular}[c]{@{}c@{}}KR\end{tabular} & \begin{tabular}[c]{@{}c@{}}19,137\\ (31.16\%)\end{tabular} & \begin{tabular}[c]{@{}c@{}}20,190\\ (32.88\%)\end{tabular} & \begin{tabular}[c]{@{}c@{}}21,844\\ (35.57\%)\end{tabular}  & \begin{tabular}[c]{@{}c@{}}241\\ (0.39\%)\end{tabular} & 61,412 \\ \hline
\end{tabular}
}
\label{sentiment} 
\end{table}

\textbf{Annotation guidelines and prompts are available in our public GitHub repository \footnote{\url{https://anonymous.4open.science/r/XXXXX-6BCB}}.}

\subsection{{Regression Analysis}}\label{regression}
Comments with annotated sentiment are used for regression analysis. We aim to predict the three sentiment types using multiple variables to examine whether one variable is associated with pro-natalist, anti-natalist, or neutral sentiment, when holding all other variables equal, aka, ceteris paribus.

We use a multivariate regression model \cite{chatfield2018introduction} and focus on whether sentiments vary based on (1) comment dimensions, (2) short video characteristics, and (3) local socioeconomic contexts, ceteris paribus. To predict the level of support for reproduction, we employed a \textit{Multinomial logistic regression} \cite{agresti2013categorical} that is suitable for predicting a categorical variable. Our dependent variable is a three-level variable (against-reproduction, neutral, and pro-reproduction), without assuming an ordered structure. The model specification is set as below:

\begin{equation}
\ln \left( \frac{P(Y_i = k)}{P(Y_i = \text{Base})} \right) = \beta_{0k} + \sum_{j=1}^{p}\beta_{jk}X_{ji} + \epsilon_i
\end{equation}

$P(Y_i = k)$ is the probability of comment \textit{i} being assigned a sentiment category \textit{k}. \textit{Base} is the neutral sentiment. Therefore, $\ln \left( \frac{P(Y_i = k)}{P(Y_i = \text{Base})} \right)$ is the logged odds of the comment being classified into category \textit{k} rather than `neutral.' $X_{ji}$ includes all the predictors, such as short-video characteristics or regional socioeconomic context (to be introduced in more details in the Results section). $\epsilon_i$ is the random error term.

While an ordinal structure (anti-natalis $<$ neutral $<$ pro-natalist) might theoretically apply to this sentiment category, a likelihood-ratio test (LR $\chi^2$(22) = 3832.66, p = 0.000) rejected the proportional odds assumption of ordered logistic regression, indicating that the multinomial model provides a statistically superior fit. 

\section{RESULTS}

\subsection{Content Dimensions and Topic Themes}

\begin{figure}[htp]
    \centering
    \includegraphics[width=8.5cm]{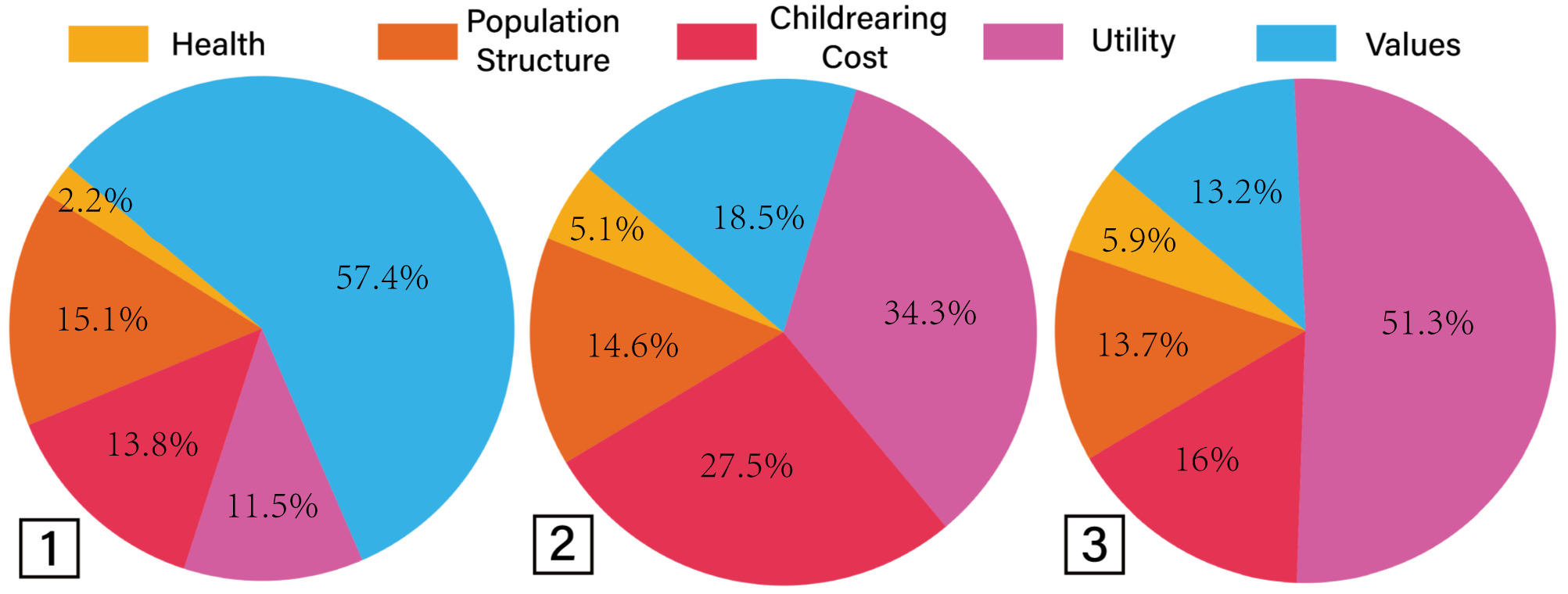}
    \caption{Distribution of the five theoretical dimensions in (1) China; (2) Japan; (3) South Korean.}
    \label{topic-1-2}
\end{figure}

Figure \ref{topic-1-2} reports the distribution of the five {dimensions} in the comments. Over 57\% of Chinese comments are about ``Values,'', while those in Japan and South Korea concentrate on ``Utility of Childbirth.'' ``Child-rearing cost'' is a common dimension in all countries. 

\begin{figure}[htp]
    \centering
    \includegraphics[width=8.5cm]{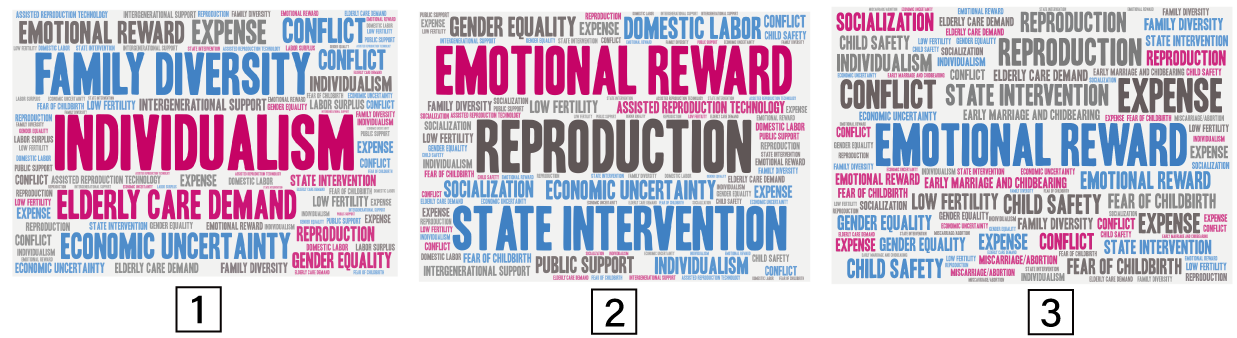} 
    \caption{Word cloud map of (1) 17 themes in Chinese Comments; (2) 18 themes in Japanese Comments; (3) 16 themes in South Korean Comments.}
    \label{topic-2}
\end{figure}

Figure \ref{topic-2} presents the word clouds featuring 21 {themes} from the word cluster grouping. The visualization is based on the words' higher weighted log odds~\cite{monroe2008fightin}, {which compares a word’s relative frequency in the target corpus to its frequency in all other comments—after a small Bayesian prior is applied—so that the words with the largest positive scores are those most distinctive of that corpus. {The exact percentages of topic themes for each country are in Appendix Table \ref{tab:theme_distribution_country}.}}

\subsubsection{Childrearing Cost Dimension}
Child-rearing costs are the most common dimension among Korean comments. 5 topic themes are included: 1. ``Expense,'' 2. ``Economic Uncertainty,'' 3. ``Domestic Labor,'' 4. ``Public Support,'' and 5. ``Child Safety''. In South Korea, ``Expense'' is the most common topic, with discussions mainly about the cost of raising children. Previous research shows that South Korea has some of the highest child-rearing costs in the world \cite{RN1}. In China, the topic ``Economic Uncertainty'' is the most common, with representative comments like \textit{``Wages will not be increasing, so I don't dare to have children"} and \textit{``I'm afraid of being laid off.''} In Japan, \textit{``Domestic Labor''} is the most frequent topic, with conversations centered around women performing housework, such as \textit{``I will be a housewife,''} \textit{``Her muscles are strong; she does housework fast.''}, followed by concerns about expenses.

\subsubsection{Utility Dimension}
Discussions about the important functions of having children are more common in Japan and South Korea. Four topic themes are connected to this dimension: 1. ``Reproduction,'' 2. ``Emotional Reward,'' 3. ``Intergenerational Support,'' 4. ``Socialization.'' In China and Japan, the topic ``Reproduction'' is the most common, with discussions focusing on having more children. Representative comments include \textit{``I love children, just had my third.''} Comments like \textit{``Continuing the family line.''} are more common in China. In Japan and South Korea, ``Emotional Reward'' is also one of the most common topics, with discussions like \textit{``Such a cute child''} and \textit{``I'm so happy to have such an adorable child.''}

\subsubsection{Values Dimension} 

This dimension has four topic themes: 1. ``Individualism,'' 2. ``Family Diversity,'' 3. ``Gender Equality,'' 4. ``Conflict.'' In China, ``Individualism'' is the most common topic, typically focused on personal freedom and hedonistic reasoning. Comments often express individualistic viewpoints such as \textit{``Only by not having children can you enjoy life,'' ``No kids mean freedom,''} and \textit{``Living her/his life well.''} In Japan, ``Gender Equality'' is the most prevalent topic, with discussions often centered on gender roles, for example, \textit{``I don't understand why childless men would criticize this. They should experience 10 months of natural childbirth or cesarean section before expressing a view.''} In South Korea, ``Conflict'' is the most common topic theme, with discussions revolving around the tension between responsibility and the pressures of childbearing, such as \textit{``I am a mother of two children. The stress of raising them has made me depressed.''} In China, expressions about the great pressure from their parents are more common. These comments reflect the tension between the more liberal, individualist views and the conventional gender or collective values embedded in parenthood and family.

\subsubsection{Population Structure Dimension}

This theoretical dimension occupies a similar proportion across the three countries and consists of five topic themes: 1. ``Low Fertility,'' 2. ``Elderly Care Demand,'' 3. ``Early Marriage and Childbearing,'' 4. ``Labor Surplus,'' and 5. ``State Intervention.'' In China, the ``Elderly Care Demand'' is the most common theme, {with its representative comment} such as \textit{``When I'm old, I'll take care of myself''} or \textit{``I’d rather enjoy now (not having kids) and face the bad consequences when older.''} In Japan and South Korea, ``State Intervention'' theme is more prevalent. Representative comments in Japan include \textit{``raising taxes''} and \textit{``government's measures against low birth rates,''} while in Korea, comments often mention \textit{``immigration to increase population''} and \textit{``government policies.''} The difference in the distribution of the themes highlights the differing focuses of public discourse in different countries: Chinese users expressed great concern for elderly care, while Japan and Korea emphasize government interventions to address the issue of shrinking population.

\subsubsection{Health Dimension}

The ``health'' dimension is the least mentioned and contains three topic themes: 1. Assisted Reproduction Technology, 2. Fear of Childbirth, and 3. Miscarriage/Abortion. In China and Japan, ``Assisted Reproduction Technology'' is the most common theme in this dimension. 
{An illustrative Chinese comment noting that \textit{``Traditional Chinese medicine helped me have a child.''}} Japanese comments include phrases like \textit{``I had a child after continued treatment.''} In Korea, ``Fear of Childbirth'' is the most common, with discussions focused on labor pain. These comments from China, Japan, and Korea highlight the different fertility-related health challenges.

\subsection{Sentiment Analysis Result}

We now report the distribution of the three types of sentiments toward childbirth and parenthood. {To simplify interpretation, the following discussions in this section focus solely on the proportion of anti-natalist comments, given this region's lowest-low fertility level. Please refer to (Appendix Figure \ref{Sentiment by Theoretical} C) for a report of all three sentiment categories.}

\begin{figure}[h]
    \centering
    \includegraphics[width= 8CM]{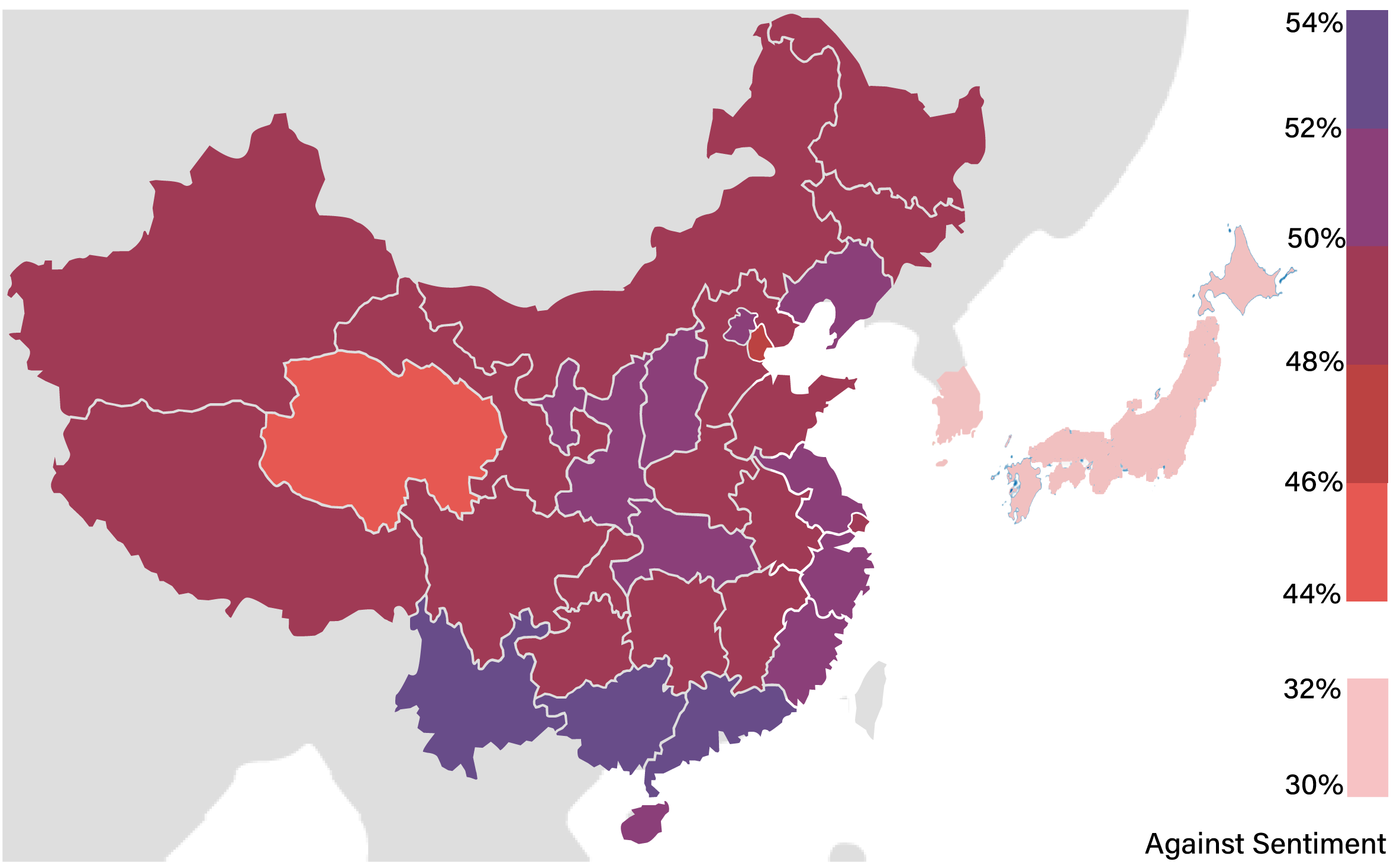}
    \caption{Levels of anti-natalist sentiment across the Chinese provinces, Japan, and South Korea.}
    \label{Sentiment-1}
\end{figure}

50.29\% of the comments in our Chinese sample were categorized as ``anti-natal.'' In the Japanese sample, this proportion was 30.11\%, and in the Korean sample, it was 31.28\%.  {Figure \ref{Sentiment-1} shows the levels of anti-natal sentiment across provinces in China, in Japan, and in South Korea.} This finding reflects a relatively stronger anti-natalism in \textit{Douyin} than in \textit{Tiktok}. 

In China, Qinghai (with a sparse population and a relatively higher level of ethnic minority groups, with Islam being the predominant religion) has the lowest anti-reproduction sentiment at 45.30\%, followed by Tianjin at 47.80\% and Chongqing at 48.13\%. These two megacities are renowned for their high economic development and relaxed lifestyles. In contrast, the southwest region exhibits higher anti-reproduction sentiment. For example, Guangxi and Yunnan provinces, with a relatively poor economic development level, show the strongest opposition to childbirth, with this sentiment at 52.52\% and 52.47\%, respectively.

\subsubsection{Anti-natal Sentiment by Content Dimensions}

\begin{figure}[h]
     \centering
     \includegraphics[width= 8.5 CM]{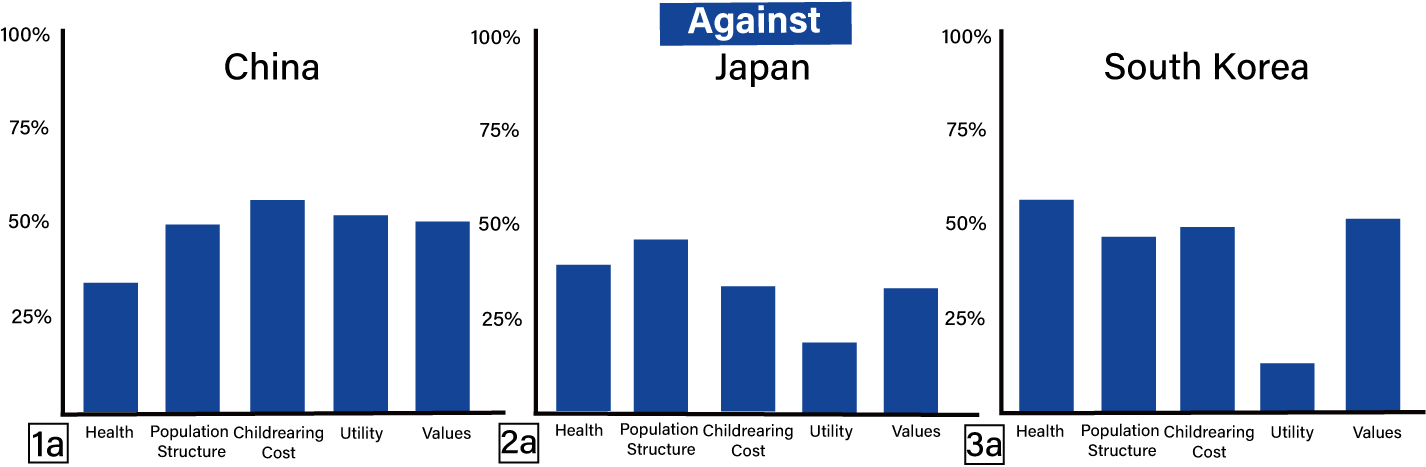}
    \caption{The distribution of anti-natalist sentiments across five dimensions in CN, JP, and SK.}
    \label{SentimentA}
 \end{figure}

{Figure \ref{SentimentA} } illustrates the distribution of the anti-natal sentiment across five dimensions. In China, the ``Childrearing Costs'' dimension generates the strongest anti-natalist comments, followed by comments related to ``Values,'' while the ``Health'' dimension generates the least anti-natalist comments. In Korea, significant opposition to childbirth is also observed in relation to childrearing costs, but is also in comments talking about ``Health.'' In Japan, the highest level of anti-natal sentiment is observed in comments related to ``Population Structure.'' When comments pertain to the 'Utility' dimension, which focuses on the role of families, attitudes are the least anti-natalist in Japan and South Korea, but this is not the case in China.

\subsubsection{Anti-natalist Sentiment by Video Characteristics} 

\begin{figure}[h]
     \centering
     \includegraphics[width= 8.5 CM]{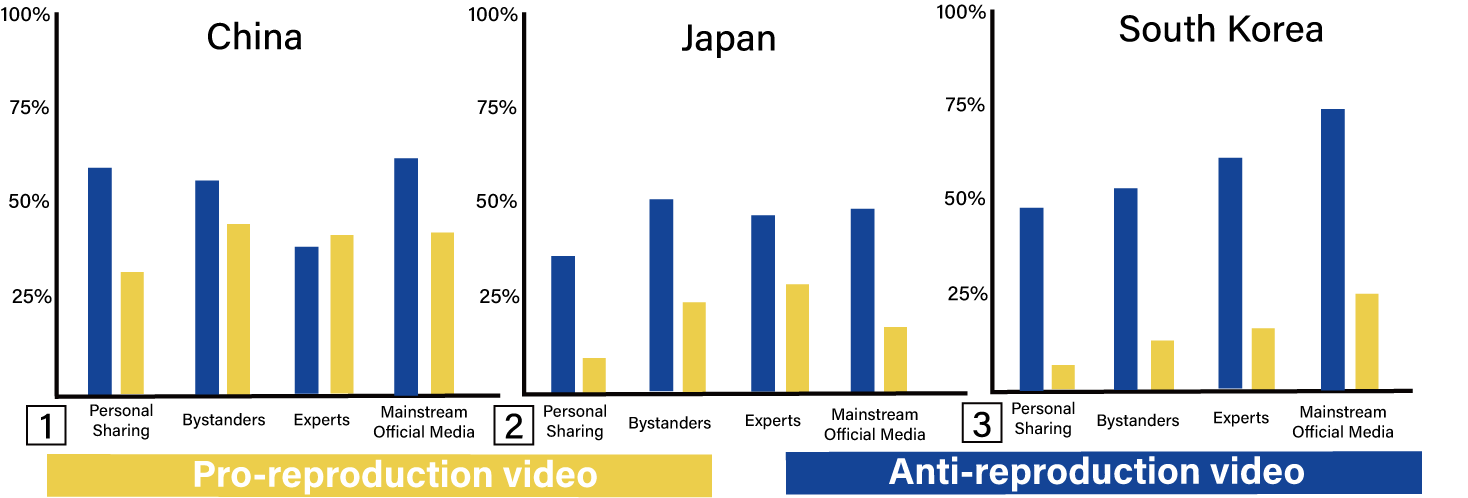}
    \caption{The distribution of anti-natalist sentiments across four video account types, further separated by video stances towards reproduction.}
    \label{SentimentB}
 \end{figure}

As mentioned above, videos are classified into four categories: personal sharing, bystanders, experts, and mainstream media; and two stances: either pro- or anti-reproduction. {Figure \ref{SentimentB} plots the proportion of anti-natalist comments across the four video categories, that is further divided by their attitudes towards reproduction. First, the proportion of anti-natalist comments is consistently higher for short videos that express an anti-reproduction stance (indicated in blue). This difference is especially significant in South Korea, followed by Japan, and then China, underscoring the ability of these short videos to influence and align attitudes. It is interesting to note that Chinese users' attitudes towards childbirth and parenthood are less influenced by the opinions presented in short videos compared to users from Japan and Korea.}

Considering the content creator's identity, Figure \ref{SentimentB} shows that mainstream accounts receive the highest number of anti-natal comments in both China and Korea, regardless of their viewpoints. In Japan, however, if the mainstream media shows a pro-natalist attitudes, the comments are much less likely to be anti-natal, showing a high level of support for pro-natalist policies {Videos that share personal experiences related to childbirth or parenthood elicit the most polarized reactions, depending on the video's stance. This highlights the strong emotional impact of personal stories.} Additionally, videos featuring experts generate fewer anti-reproduction comments in China and Japan, but not in South Korea.

\subsection{Regression Results}

{Previous bivariate analyses of how sentiment varies across different short video account types may be influenced by the varying stances of the short videos or other characteristics. We now use multivariate regressions to examine whether the sentiment still varies across (1) the theoretical dimensions of the comment, (2) the corresponding short video characteristics, (3) multiple local socioeconomic indicators (mostly for the Chinese comments with geotags), after holding everything else the same using the regression methods.} 

For the socioeconomic context where the comment comes from, as discussed in the subsection titled ``Socioeconomic Development and Responses,'' we focus on economic development, urbanization, and gender equality levels. GDP per capita\footnote{GDPP for Japan and South Korea are converted from the 2021 World Bank data reported in U.S. dollars. The exchange rate used for conversion is the average of the 2021 RMB/USD central parity rates, with a maximum of 6.3498 and a minimum of 6.5713, yielding an average rate of 6.46055.}, representing the Regional socioeconomic development level, is from China's 7th National Population Census and the World Bank\footnote{\url{https://datacatalog.worldbank.org/collections}} and the United Nations\footnote{\url{https://data.un.org/}}). Urbanization level is the percentage of the total population living in urban areas. We used the male-to-female population ratio to measure son preferences, which reflects traditional family values and gender equality levels. We included female labor force participation rates to represent the economic cost of childbirth for women. In the model, we also included a few control variables, such as the provincial total fertility rate, Internet penetration rate, and the comment year. {Detailed descriptive statistics (mean sentiment score, standard deviation, and sample size) of the variables for each country and region are presented in Appendix Table \ref{appendix:desc}. }

{We conducted country-specific regressions.} As the Japanese and South Korean datasets do not include IP location information, it was not possible to incorporate sub-national context variables in the models. Consequently, the analyses for Japan and South Korea focus solely on the relationship between sentiment and dimensions and short video characteristics. {Table ~\ref{Mlogit_country} presents the estimated coefficient of the variables, highlighting the relationship between each variable and its predicted logged odds of the specific sentiment category vs. the neutral sentiment, ceteris paribus. Appendix Table \ref{Appendix:Mlogit_pooled} presents the pooled-country sample results and shows that compared to Chinese comments, Japanese and Korean comments are more likely to be neutral. This finding is consistent with Figure \ref{Sentiment-1} .}

\begin{table}[H]
\caption{Multinomial logistic regressions predicting the logged odds of `negative' and `positive' sentiment vs. 'neutral' sentiment by country}
\scalebox{0.58}{
\begin{tabular}{lcccccc}
\hline
 & \multicolumn{2}{c}{China} & \multicolumn{2}{c}{Korea} & \multicolumn{2}{c}{Japan} \\
Base: Neutral & Anti-natal & Pro-natal & Anti-natal & Pro-natal & Anti-natal & Pro-natal \\
\hline
Dimension: Health  & -0.391*** & 0.513*** & -0.364*** & -0.193*** & 0.582*** & 1.254*** \\
vs Childrearing Cost & (0.073) & (0.074) & (0.049) & (0.058) & (0.043) & (0.045) \\
Dimension: Pop. Stru. & -0.317*** & -0.024 & -0.413*** & -0.551*** & 0.257*** & -0.536*** \\
 vs Childrearing Cost & (0.034) & (0.042) & (0.037) & (0.047) & (0.026) & (0.038) \\
Dimension: Utility  & -0.031 & 0.337*** & -1.474*** & 0.174*** & -0.276*** & 0.910*** \\
vs Childrearing Cost & (0.038) & (0.045) & (0.033) & (0.036) & (0.024) & (0.023) \\
Dimension: Values  & -0.412*** & -0.272*** & -0.395*** & -0.600*** & 0.211*** & 0.634*** \\
vs Childrearing Cost & (0.027) & (0.035) & (0.037) & (0.049) & (0.026) & (0.028) \\
\hline
Account: Bystanders  & 0.124*** & 0.204*** & 0.273*** & -0.317*** & 0.526*** & -0.083** \\
 vs Personal Sharing & (0.022) & (0.027) & (0.032) & (0.025) & (0.031) & (0.036) \\
Account: Main Media & 0.246*** & 0.339*** & 0.889*** & -0.164*** & 0.266*** & -0.180*** \\
vs Personal Sharing  & (0.046) & (0.056) & (0.038) & (0.040) & (0.024) & (0.031) \\
Account: Experts  & -0.275*** & 0.002 & 0.482*** & -0.249*** & 0.407*** & -0.031 \\
vs Personal Sharing  & (0.040) & (0.049) & (0.059) & (0.064) & (0.027) & (0.034) \\
Video attitude: Pro- & -0.271*** & 1.170*** & -1.504*** & 1.451*** & -0.920*** & 1.360*** \\
 vs Anti-reproduction & (0.023) & (0.027) & (0.026) & (0.028) & (0.022) & (0.021) \\
Video figure: Male & -0.158*** & -0.530*** & 0.042* & -0.100*** & 0.125*** & -0.036 \\
vs Female  & (0.027) & (0.032) & (0.024) & (0.022) & (0.020) & (0.023) \\
Video figure: Mid-age  & 0.136*** & 0.224*** & -0.031 & -0.069 & 0.008 & 0.011 \\
vs Youth & (0.027) & (0.033) & (0.033) & (0.042) & (0.020) & (0.019) \\
Video figure age: Old  & -0.076*** & 0.059* & -0.372*** & 0.370*** & 0.270*** & -0.047 \\
vs Youth & (0.029) & (0.035) & (0.129) & (0.083) & (0.035) & (0.049) \\
\hline
Std. GDP per capita & 0.048 & 0.109** & & & & \\
 & (0.035) & (0.045) & & & & \\
Urbanization Rate & -0.605* & -0.732* & & & & \\
 & (0.314) & (0.401) & & & & \\
 Sex Ratio & 0.648* & -0.314 & & & & \\
 & (0.349) & (0.447) & & & & \\
 Female Labor Force  & 0.113 & 0.043 & & & & \\
Participation Rate & (0.237) & (0.301) & & & & \\
Total Fertility Rate & -0.091* & -0.018 & & & & \\
 & (0.051) & (0.066) & & & & \\
Internet Penetration \% & 0.312 & 0.445 & & & & \\
 & (0.296) & (0.376) & & & & \\
Constant & -0.220 & -0.560 & 0.690*** & -0.887*** & -0.177*** & -1.754*** \\
 & (0.528) & (0.674) & (0.074) & (0.057) & (0.040) & (0.045) \\
 \hline
Year Fixed Effect & Yes & & Yes & & Yes & \\
\hline
R2 & 0.0554 & & 0.2169 & & 0.1317 & \\
N & 72,595 & & 61,171 & & 82,670 & \\
\hline
\multicolumn{7}{l}{\footnotesize * p\textless{}0.1 \quad ** p\textless{}0.05 \quad *** p\textless{}0.01} \\
\end{tabular}}
\label{Mlogit_country}
\end{table}

\subsubsection{{Sentiment and Content Dimensions}}
We first examine how different discussion dimensions of the comments would reflect different sentiments. Compared to comments discussing ``childbearing cost,'' comments about ``Health'' are more likely to be pro-natal in China, more neutral in Korea, and less neutral in Japan. Compared to those discussing ``childrearing costs,'' Chinese comments discussing the ``population structure'' are much less likely to be anti-natal, highlighting their great concerns over old age care. Korean comments falling into this dimension tend to be more neutral, while Japanese comments are more anti-natal and less pro-natal vs. neutral. Comments about ``Utility'' consistently express pro-natalist sentiment in all countries. Comments about ``Values'' are more likely to express neutral sentiments in China and Korea, while they tend to express more conflicting sentiments in Japan.

\subsubsection{{Sentiment and Video Characteristics}}

Compared to comments on personal-sharing videos, comments on bystander videos in China are more likely to express anti-natalist or pro-natalist sentiments rather than neutral sentiments. In Japan and South Korea, bystander videos are more likely to provoke anti-natalist sentiments compared to neutral ones, and they are less likely to elicit pro-natalist sentiments. In all countries, mainstream media are more likely to provoke anti-natalist views compared to neutral views; however, in China, mainstream media are also more likely to generate pro-natalist views. Overall, everything else being the same, the bystander and mainstream media in China seem to elicit more polarized comments but in Korea and Japan, they tend to provoke stronger anti-natal sentiment.

In all countries, short videos that express pro-natalist views are more likely to elicit pro-natalist sentiments and less likely to induce anti-natalist sentiments compared to neutral ones.

Regarding the main figures featured in the short videos, those that include male figures are less likely to provoke either anti-natalist or pro-natalist comments in China, but they are more likely to induce anti-natalist comments in Korea and Japan.

\subsubsection{{Sentiment and Regional Contexts}}
Users' attitudes could be affected by the level of economic development, urbanization, and gender equality. GDPP is positively correlated with pro-natalist sentiment ($\beta $ = 0.109**). This finding supports that higher economic development reduces the financial pressures of childbearing, such as through higher incomes, a more prosperous labor market, and an overall optimistic outlook for the future. 

 Urbanization rate is correlated with more neutral sentiment. On the one hand, high housing prices brought by urbanization would suppress fertility desire. On the other hand, urban areas offer better childcare services, education, and healthcare.
 
 The positive coefficient of the male-to-female sex ratio in predicting the logged odds of anti-natal vs. neutral sentiment ($\beta $ = 0.648*) suggests that as the number of males exceeds the number of females to a larger extent, comments become more anti-natal. Female labor force participation rates are not found to be statistically significantly correlated with comment sentiments. Comments from regions with a higher fertility level tend to be less likely to reveal an anti-natalist sentiment  ($\beta $ = -0.091*).

{We conducted the same analysis using the pooled Chinese, Japanese, and Korean samples with Japanese and Korean national statistics (more details in Appendix Table \ref{Appendix:Mlogit_pooled}). The findings related to these regional contexts remain unchanged.}

\section{DISCUSSIONS}
Fertility rates in East Asia continue to be among the lowest in the world, {despite the pro-natalist programs introduced in Japan and South Korea since the 1990s, as well as China's recent shift away from its one-child policy between 2016 and 2021 \cite{yang2022china}.} The widespread algorithmic feeds of short-form videos, coupled with the growing uncertainties of the economic future, imply a pessimistic perspective of childbirth and parenthood that young users engage with. Investigating whether and how this pessimistic picture spreads online is crucial for uncovering the deeper reasons behind this extremely low fertility trend. In this paper, we analyzed short videos and social media comments discussing childbirth and parenthood on popular platforms. We reported the main topics and embedded sentiments of these comments, how these sentiments were driven by various short video characteristics, as well as what regional context may shape these sentiments in three East Asian countries.

\subsection{Online Discussion Dimensions and Themes} 

Three dimensions--\textit{Childrearing Costs}, \textit{Utility of children}, and \textit{Values}--dominate online discussions. Particularly in China and South Korea, high expenses related to housing and education, along with significant dissatisfaction with income growth and savings, pose major barriers to family formation. The perceived utility of children varies by culture. In China and Japan, children are essential for continuing the family line. In contrast, South Korea places more emphasis on the emotional rewards. {``Values'' occupy the central dimension in China with a strong emphasis on individual freedom, with 30\% of the comments belonging to the individualism theme. Typical phrases include \textit{``Only by not having children can you enjoy life.''} or \textit{``Enjoy life.''} } In Japan and South Korea, discussions concentrate on the conflict between traditional gender roles and women's self-actualization. 

\subsection{Online Discussion Sentiments} 
In China, attitudes towards childbirth and parenthood are predominantly negative, primarily driven by economic concerns and individualism. In contrast, Japan and South Korea exhibit more balanced views, represented by a high level of comments with neutral sentiment. 

Comments discussing ``Utility of children'' express more pro-natalist sentiments in all countries. However, the relationships between comments discussing other dimensions and their sentiment differ across countries, uncovering regional heterogeneity in the more specific topics related to childbirth and parenthood, as presented in Figures 2 and 3.

\subsubsection{Short Video and Response Sentiments} 
The sentiment expressed in comments is also strongly influenced by video characteristics. Comment sentiments align closely with the attitudes expressed in short videos. Regression results show that compared to personal sharing, the bystander and mainstream media videos in China seem to elicit more conflicting sentiments. However, in Korea and Japan, they tend to provoke stronger anti-natal sentiment. This is likely due to perceived disconnects between official messages and viewers' realities. The gender and age of the main figure displayed in the short videos also influence the comment sentiments.

\subsubsection{Regional Sentiment Dynamics} 
Local socioeconomic context is associated with the attitudes expressed online. Economic development level, represented by GDP per capita, tends to foster more pro-natal sentiment. Areas with lower economic development, like the Southwest provinces in China, exhibit stronger anti-natalist sentiments, where financial insecurity and economic pressures dominate the discourse. Regression results also suggest that comments from regions with strong son preferences demonstrate a higher level of anti-natalism. Moreover, comments from regions with a relatively high fertility level tend to be less likely to express negative sentiment towards childbirth and parenthood, highlighting the significance of fertility norms.

\subsection{Limitations and Future Work}
This study has several limitations, mainly related to the use of social media data. The data is drawn from users who actively post comments, which may not accurately represent the broader user base or the general population~\cite{inara2021let}. Those who express their views online may tend to have more liberal, individualistic attitudes, potentially skewing our conclusions toward a more pessimistic view of reproduction. Therefore, the findings reflect only the content publicly discussed on these platforms and should not be generalized to the overall population in these countries. However, we still consider this analysis important as the information left by potentially highly selective user groups remains to be consumed by all users on social media platforms and can have a great impact on the general public who spend increasingly more time online. {Additionally, we lack information about the commenters' personal details, such as their gender and age, apart from their provincial-level geolocation in China. This limits our ability to analyze how gender and age dynamics influence the discourse. However, according to the \textit{TikTok} business report, we know that the majority of users are young and of prime childbearing age.} The second limitation is that the IP geolocation data on \textit{Douyin} is limited to the provincial level within China. Considering the huge population size of each province, subtle regional differences that play a role in shaping public discourse might be overlooked. Finally, while the use of LLMs offers more nuanced labels with the complex sentiments embedded in these discussions, LLMs are known to produce different results each time they perform the task, which can hinder replication. Notably, the same issue exists for human annotators, and the performance stability of the LLM, in terms of accuracy level, is good in our case. {Nevertheless, although LLM accuracy is high and comparable across the three languages, residual language-specific bias cannot be entirely excluded.}

\section{CONCLUSIONS}
Online discussions about childbirth and parenthood focus on childrearing costs, the utility of children, and tensions between individualistic values and traditional family and gender norms. Notable country differences emerge, with comments from China reflecting the highest levels of anti-natalism and significant concerns about income and child-rearing costs. Views in Japan and South Korea are more balanced. Additionally, sentiment in comments is influenced by various short video characteristics, with its stance posing the strongest impact. Comments from areas with higher economic development and gender equality also show more neutral or pro-natalist sentiments. These insights offer an important contribution to the field of social media analysis, particularly in understanding the formulation and spreading of views towards childbirth and parenthood on an algorithmic-led video sharing platform.

\textbf{ETHICAL STATEMENT}
Data is from publicly available platforms. We removed user names and stored the link to short videos separately. Regarding the provincial-level geo-location information for Chinese comments, given that even the smallest province has at least one million people, it is unlikely that the universal disclosure of users' IP geo-location will enable the identification users. All data remained anonymous with a 10\% random sample published in our repository with three variables--comments, topic label, and sentiment label. This project is approved by our Institutional Review Board (IRB XXXX).

\bibliography{ICWSM}

\clearpage
\appendix
\section{APPENDIX}

\subsection{Hashtag Keywords}\label{appendix:A1}
\begin{table}[ht]
\caption{60 Hashtag Keywords sets by country}
\scalebox{0.7}{
\begin{tabular}{ccc}
\hline
\textbf{China (Douyin)}                                                                   & \textbf{Japan (TikTok)}                                                              &  \textbf{South Korea (TikTok) }                                                          \\ \hline
DINK                                                                                      & Single for Life                                                                      & \begin{tabular}[c]{@{}c@{}}Childbirth Incentive \\ Policies\end{tabular}       \\
\begin{tabular}[c]{@{}c@{}}Not Having \\ Children\end{tabular}                            & Single-Minded                                                                        & Childbirth Subsidies                                                           \\
\begin{tabular}[c]{@{}c@{}}Not Getting \\ Married and Not \\ Having Children\end{tabular} & Childless Couple                                                                     & \begin{tabular}[c]{@{}c@{}}Childcare Support \\ System\end{tabular}            \\
Non-Married                                                                               & Childless Winners                                                                    & Happy Parenting                                                                \\
No Children                                                                               & Childless Losers                                                                     & \begin{tabular}[c]{@{}c@{}}Overcome Low \\ Birth Rates\end{tabular}            \\
\begin{tabular}[c]{@{}c@{}}Young People \\ Don't Want to \\ Have Children\end{tabular}    & \begin{tabular}[c]{@{}c@{}}Don't Want \\ Children\end{tabular}                       & Have a Child Early                                                             \\
Have a Child Early                                                                        & No Children                                                                          & High School Mums                                                               \\
Single Parents                                                                            & \begin{tabular}[c]{@{}c@{}}No Children \\ for Life\end{tabular}                      & \begin{tabular}[c]{@{}c@{}}Multi-generational \\ Families\end{tabular}         \\
\begin{tabular}[c]{@{}c@{}}Not Having \\ Children to \\ Enjoy Life\end{tabular}           & \begin{tabular}[c]{@{}c@{}}Having Children is \\ Difficult\end{tabular}              & \begin{tabular}[c]{@{}c@{}}The Era of Ultra-low \\ Birth Rates\end{tabular}    \\
Married and Sterile                                                                       & Low Birthrate                                                                        & Declining Birth Rate                                                           \\
\begin{tabular}[c]{@{}c@{}}Many Children, \\ More Blessings\end{tabular}                       & \begin{tabular}[c]{@{}c@{}}Measures to Combat \\ the Low Birthrate\end{tabular}      & Childcare Stress                                                               \\
Family Prosperity                                                                         & \begin{tabular}[c]{@{}c@{}}Support for \\ Raising Children\end{tabular}              & Childcare Costs                                                                \\
\begin{tabular}[c]{@{}c@{}}Have More \\ Children\end{tabular}                             & Large Family                                                                         & Parenting Hell                                                                 \\
\begin{tabular}[c]{@{}c@{}}Have Both a Son \\ and a Daughter\end{tabular}                 & \begin{tabular}[c]{@{}c@{}}Large Family \\ Mother\end{tabular}                       & Solo Parenting                                                                 \\
\begin{tabular}[c]{@{}c@{}}Have Children \\ Early\end{tabular}                            & Many Children                                                                        & \begin{tabular}[c]{@{}c@{}}Decreased Willingness \\ to Give Birth\end{tabular} \\
Three Children                                                                            & \begin{tabular}[c]{@{}c@{}}Many children, with \\ Children Close in Age\end{tabular} & \begin{tabular}[c]{@{}c@{}}Not Having \\ Children\end{tabular}                 \\
Childcare Costs                                                                           & Want Children                                                                        & \begin{tabular}[c]{@{}c@{}}Reasons for Not \\ Having Children\end{tabular}     \\
\begin{tabular}[c]{@{}c@{}}Encourage \\ Childbirth\end{tabular}                           & \begin{tabular}[c]{@{}c@{}}Desired Number \\ of Children\end{tabular}                & Dink Family                                                                    \\
Childbirth Subsidy                                                                        & Lovely Child                                                                         & Single People                                                                  \\
Single People                                                                             & Happy Parenting                                                                      & Non-Married                                                                    \\ \hline
\end{tabular}
}
\label{A1}
\end{table}

\subsection{More Detailed Distribution of Topic Analysis Results}\label{appendix:A2-3}

\begin{table}[H]
\centering
\caption{Distribution of topic themes by country}
\scalebox{0.67}{
\begin{tabular}{llccc}
\hline
\textbf{Dimensions} & \textbf{Topic Themes} & \textbf{China} & \textbf{Japan} & \textbf{Korea} \\
\hline
Childrearing  &  & 13.76 & 27.01 & 16.00 \\
Cost & Child Safety & 0.00 & 1.21 & 7.52 \\
 & Domestic Labor & 1.08 & 8.87 & 0.00 \\
 & Economic Uncertainty & 6.62 & 7.71 & 0.55 \\
 & Expense & 5.87 & 3.26 & 7.93 \\
 & Public Support & 0.20 & 5.95 & 0.00 \\
\hline
Utility &  & 11.51 & 34.51 & 51.27 \\
 & Emotional Support & 5.09 & 12.33 & 35.21 \\
 & Intergenerational Support & 2.99 & 2.18 & 0.00 \\
 & Reproduction & 3.43 & 14.10 & 14.53 \\
 & Socialization & 0.00 & 5.90 & 1.53 \\
\hline
Values &  & 57.37 & 18.64 & 13.16 \\
 & Conflict & 6.12 & 2.02 & 5.24 \\
 & Family Diversity & 17.45 & 2.12 & 2.84 \\
 & Gender Equality & 3.84 & 8.28 & 3.58 \\
 & Individualism & 29.97 & 6.22 & 1.51 \\
\hline
Population  &  & 15.12 & 14.74 & 13.68 \\
Structure & Early Marriage and Childbearing & 0.00 & 0.00 & 2.34 \\
 & Labor Surplus & 1.89 & 0.00 & 0.00 \\
 & Low Fertility & 2.29 & 3.41 & 0.74 \\
 & Old Age Care Demand & 8.27 & 1.16 & 1.58 \\
 & State Intervention & 2.68 & 10.17 & 9.02 \\
\hline
Health &  & 2.24 & 5.10 & 5.89 \\
 & Assisted Reproduction & 1.36 & 3.18 & 0.00 \\
 & Fear of Childbirth & 0.88 & 1.92 & 4.89 \\
 & Miscarriage & 0.00 & 0.00 & 1.01 \\
\hline
\textbf{Total} & & \textbf{100.00} & \textbf{100.00} & \textbf{100.00} \\
\hline
\end{tabular}
}
\label{tab:theme_distribution_country}
\end{table}

\begin{table}[ht]
\centering
\caption{Distribution of the five dimensions by country}
\scalebox{0.7}{
\begin{tabular}{ccccccc}
\hline
                                                       & Health                                                  & \begin{tabular}[c]{@{}c@{}}Population \\ Structure\end{tabular} & \begin{tabular}[c]{@{}c@{}}Childrearing\\ Cost\end{tabular} & Utility                                                   & Values                                                    & Total \\ \hline
CN                                                  & \begin{tabular}[c]{@{}c@{}}1,675\\ (2.24\%)\end{tabular} & \begin{tabular}[c]{@{}c@{}}11,316\\ (15.12\%)\end{tabular}       & \begin{tabular}[c]{@{}c@{}}10,296\\ (13.76\%)\end{tabular}   & \begin{tabular}[c]{@{}c@{}}8,613\\ (11.51\%)\end{tabular}  & \begin{tabular}[c]{@{}c@{}}42,928\\ (57.37\%)\end{tabular} & 74,828 \\
JP                                                  & \begin{tabular}[c]{@{}c@{}}4,227\\ (5.10\%)\end{tabular} & \begin{tabular}[c]{@{}c@{}}12,214\\ (14.74\%)\end{tabular}       & \begin{tabular}[c]{@{}c@{}}22,390\\ (27.01\%)\end{tabular}   & \begin{tabular}[c]{@{}c@{}}28,604\\ (34.51\%)\end{tabular} & \begin{tabular}[c]{@{}c@{}}15,452\\ (18.64\%)\end{tabular} & 82,887 \\
\begin{tabular}[c]{@{}c@{}}SK \end{tabular} & \begin{tabular}[c]{@{}c@{}}3,619\\ (5.89\%)\end{tabular} & \begin{tabular}[c]{@{}c@{}}8,399\\ (13.68\%)\end{tabular}        & \begin{tabular}[c]{@{}c@{}}9,824\\ (16.00\%)\end{tabular}    & \begin{tabular}[c]{@{}c@{}}31,487\\ (51.27\%)\end{tabular} & \begin{tabular}[c]{@{}c@{}}8,083\\ (13.16\%)\end{tabular}  & 61,412 \\ \hline
\end{tabular}
}
\label{topic-1}
\end{table}

\subsection{{Check whether the distribution of short videos is sensitive to the crawling duration}}

We selected \textit{TikTok} samples from Japan (JP) and South Korea (SK) that were crawled for the same 15-day duration as the China (CN) \textit{Douyin} sample and prepared a distribution table below. The characteristics of these reduced short video samples are consistent with those in the full sample in Table 2. Given the consistent video features across different crawling time durations, there is no evidence to suggest that the extended crawling of the JP and SK data with more short video numbers would bias the results.

\begin{table}[h]
\centering
\caption{Distribution of short video characteristics of the \textbf{reduced} JP and SK samples}
\begin{tabular}{ccccc}
\hline
Video Feature                                                              & JP     & Total                & SK     & Total                \\ \hline
\begin{tabular}[c]{@{}c@{}}Video stance- \\ Anti-reproduction\end{tabular} & 53.5\% & \multirow{2}{*}{275} & 45\%   & \multirow{2}{*}{200} \\
Pro-reproduction                                                           & 46.5\% &                      & 55\%   &                      \\ \hline
Gender-M                                                                   & 35.6\% & \multirow{2}{*}{275} & 36\%   & \multirow{2}{*}{200} \\
Gender-F                                                                   & 64.4\% &                      & 64\%   &                      \\ \hline
Age-Young                                                                  & 43.6\% & \multirow{3}{*}{275} & 85\%   & \multirow{3}{*}{200} \\
Age-Middle                                                                 & 51.6\% &                      & 14\%   &                      \\
Age-Old                                                                    & 4.7\%  &                      & 1\%    &                      \\ \hline
Personal Sharing                                                           & 58.9\% & \multirow{4}{*}{275} & 38.5\% & \multirow{4}{*}{200} \\
Bystanders                                                                 & 6.5\%  &                      & 41.5\% &                      \\
Experts                                                                    & 16.7\% &                      & 15\%   &                      \\
Mainstream Media                                                           & 17.8\% &                      & 5\%    &                      \\ \hline
\end{tabular}
\label{Appendix: reducedvideosample}
\end{table}

\subsection{LLM sentiment annotation performance: [1] Trial annotation}\label{appendix: A2-3}

\begin{table}[ht]
\centering
\caption{Trial sentiment distribution: same 1,000 comments, human vs. \textit{Qwen} annotation.}
\scalebox{1}{
\begin{tabular}{ccccc}
\hline
Country                                                                & Sentiment & \begin{tabular}[c]{@{}c@{}}Manual \\ labels\end{tabular} & \begin{tabular}[c]{@{}c@{}}Tongyi\\ Qwen\end{tabular} & Accuracy                \\ \hline
\multirow{3}{*}{China}                                                 & Anti-natal   & 569                                                      & 518                                                      & \multirow{3}{*}{80.8\%} \\
                                                                       & Pro-natal   & 140                                                      & 187                                                      &                         \\
                                                                       & Neutral   & 291                                                      & 295                                                      &                         \\
\multirow{3}{*}{Japan}                                                 & Anti-natal   & 296                                                      & 303                                                      & \multirow{3}{*}{92.9\%} \\
                                                                       & Pro-natal   & 283                                                      & 282                                                      &                         \\
                                                                       & Neutral   & 421                                                      & 415                                                      &                         \\
\multirow{3}{*}{\begin{tabular}[c]{@{}c@{}}South\\ Korea\end{tabular}} & Anti-natal   & 365                                                      & 339                                                      & \multirow{3}{*}{93.3\%} \\
                                                                       & Pro-natal   & 271                                                      & 298                                                      &                         \\
                                                                       & Neutral   & 364                                                      & 363                                                      &                         \\ \hline
\end{tabular}
}
\label{sentiment-test} 
\end{table}

\subsection{LLM sentiment annotation performance: [2] Full scale annotation.} \label{appendix: A3-4}

\begin{table}[ht]
\centering
\caption{1,000 random samples from the full-scale LLM annotation for each country.}
\scalebox{0.8}{
\begin{tabular}{ccccc}
\hline
Country                                                                 & Sentiment   & \begin{tabular}[c]{@{}c@{}}Manual \\ Labels\end{tabular} & \begin{tabular}[c]{@{}c@{}}Tongyi\\ Qwen\end{tabular} & Accuracy                \\ \hline
\multirow{4}{*}{China}                                                  & Anti-natal      & 478                                                      & 498                                                      & \multirow{4}{*}{88.4\%} \\
                                                                        & Pro-natal      & 203                                                      & 210                                                      &                         \\
                                                                        & Neutral      & 316                                                      & 287                                                      &                         \\
                                                                        & Unidentified & 3                                                        & 4                                                        &                         \\
\multirow{4}{*}{Japan}                                                  & Anti-natal      & 253                                                      & 291                                                      & \multirow{4}{*}{90.2\%} \\
                                                                        & Pro-natal      & 296                                                      & 291                                                      &                         \\
                                                                        & Neutral      & 449                                                      & 415                                                      &                         \\
                                                                        & Unidentified & 2                                                        & 3                                                        &                         \\
\multirow{4}{*}{\begin{tabular}[c]{@{}c@{}}South \\ Korea\end{tabular}} & Anti-natal      & 334                                                      & 331                                                      & \multirow{4}{*}{97.7\%} \\
                                                                        & Pro-natal      & 316                                                      & 319                                                      &                         \\
                                                                        & Neutral      & 345                                                      & 345                                                      &                         \\
                                                                        & Unidentified & 5                                                        & 5                                                        &                         \\ \hline
\end{tabular}
}
\label{sentiment-test2}
\end{table}

\subsection{Distribution of the three sentiment categories}

\begin{figure}[ht]
    \centering
    \includegraphics[width=7.5CM]{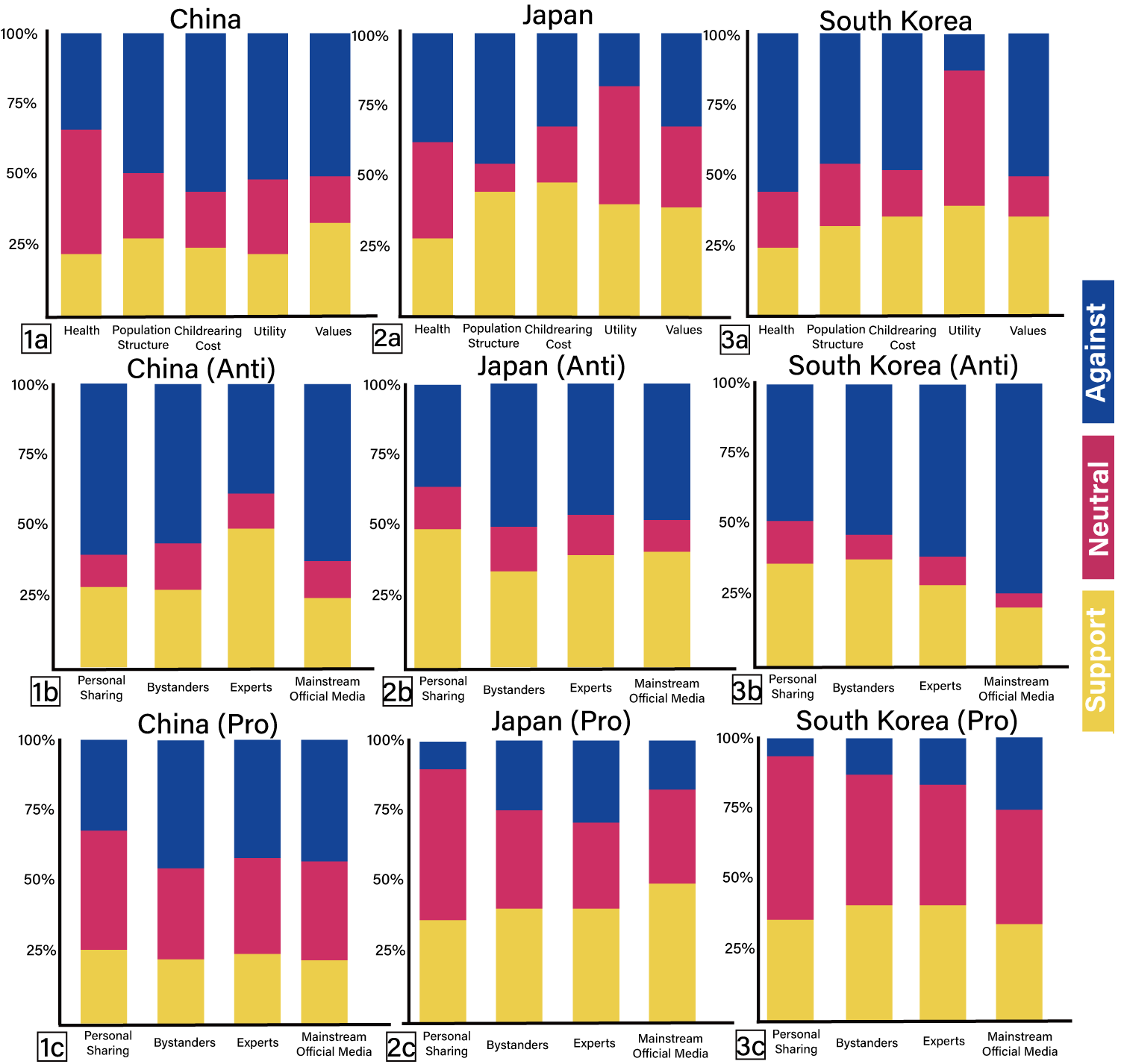}
    \caption{The distribution of sentiment categories across (1a)(2a)(3a) five dimensions, (1b)(2b)(3b) four video account types with an anti-reproduction stance, and (1c)(2c)(3c) four video account types with a pro-reproduction stance.}
    \label{Sentiment by Theoretical}
\end{figure}

\subsection{Regression with pooled sample}
{We pooled comments from China, Japan, and South Korea together. To prevent the results from being skewed by the large number of comments from JP and SK that are not proportional to their population sizes vs. China, we randomly sampled comments from JP and SK to create a dataset with the number of comments that is proportional to each country's population size. Specifically, we used Guangdong Province, which has a population nearly equal to that of Japan, and Zhejiang Province, which has a population size similar to South Korea, as references to draw random samples from JP and SK. We also dropped five provinces with less than 1\% of the comments, namely Qinghai, Tibet, Hainan, Ningxia, and Gansu. The final regression sample size was 86,508, holding the consistent proportions of video types, sub-themes, and sentiment scores as in full sample.}

{The results show that, everything else being the same, compared to Chinese comments, those Korean and Japanese comments are less likely to express either anti- or pro-natal sentiment. Overall, Japanese and Korean comments tend to reveal a more neutral sentiment. The conclusion related to the regional socioeconomic indicators remains the same as for the country-specific regression results.}

\begin{table}[ht]
\caption{Multinomial logistic model predicting the logged odds of `anti-natal' and `pro-natal' sentiments vs. `neutral' sentiment of the Pooled sample}
\scalebox{0.8}{
\begin{tabular}{lcc}
\hline
 & \multicolumn{2}{c}{Pooled country data} \\
Basement Sentiment: Neutral & Anti-natal & Pro-natal \\
\hline
 Country: South Korea vs China & -0.904*** & -0.685*** \\
 & (0.111) & (0.137) \\
Country: Japan vs China & -0.882*** & -0.632*** \\
 & (0.129) & (0.163) \\
 \hline
 Dimension: Health vs Cost & -0.057 & 0.659*** \\
 & (0.059) & (0.061) \\
Dimension: Population Structure vs Cost & -0.245*** & -0.033 \\
 & (0.029) & (0.037) \\
Dimension: Utility vs Cost & -0.184*** & 0.476*** \\
 & (0.030) & (0.035) \\
Dimension: Values vs Cost & -0.359*** & -0.154*** \\
 & (0.024) & (0.031) \\
 \hline
Account: Bystanders vs Personal Sharing & 0.164*** & 0.114*** \\
 & (0.020) & (0.025) \\
Account: Main Media vs Personal Sharing & 0.332*** & 0.048 \\
 & (0.033) & (0.042) \\
Account: Experts vs Personal Sharing & -0.163*** & -0.130*** \\
 & (0.032) & (0.039) \\
Video stance: Pro- vs Anti-reproduction & -0.430*** & 1.237*** \\
 & (0.019) & (0.022) \\
Video figure gender: Male vs Female & -0.034 & -0.404*** \\
 & (0.021) & (0.026) \\
Video figure age: Middle-aged vs Youth & 0.049** & 0.076*** \\
 & (0.022) & (0.027) \\
Video figure age: Old vs Youth & 0.010 & -0.027 \\
 & (0.026) & (0.031) \\
 \hline
Std. GDP per capita & 0.051 & 0.098** \\
 & (0.035) & (0.045) \\
Urbanization Rate & -0.619** & -0.631 \\
 & (0.313) & (0.401) \\
Internet Penetration Rate & 0.326 & 0.490 \\
 & (0.296) & (0.376) \\
Sex Ratio & 0.648* & -0.421 \\
 & (0.348) & (0.447) \\
Total Fertility Rate & -0.088* & -0.012 \\
 & (0.051) & (0.066) \\
Female Labor Force Participation Rate & 0.086 & 0.051 \\
 & (0.237) & (0.301) \\
Constant & 0.555 & -0.272 \\
 & (0.531) & (0.677) \\ \hline
Year Fixed Effect &\multicolumn{2}{c}{Yes}\\
\hline
R2 & \multicolumn{2}{c}{0.0725}  \\
N & \multicolumn{2}{c}{86,508} \\
\hline
\multicolumn{3}{l}{\footnotesize * p\textless{}0.1  ** p\textless{}0.05  *** p\textless{}0.01} \\
\end{tabular}}
\label{Appendix:Mlogit_pooled}
\end{table}

\subsection{Sample descriptive statistics by country}
\begin{table}[H]
\centering
\caption{Descriptive statistics by country}
\scalebox{0.65}{
\begin{tabular}{p{2.5cm}rccccc}
\hline
 & JP & SK & CN &  &  &  \\
\hline
\textbf{Regional context variables} & Mean& Mean& Mean& Std & Min & Max  \\
Std. GDP per capita 2021 & 3.436 & 2.819 & -0.053 & 0.590 & -0.873 & 1.619 \\
Urbanization rate 2021 & 0.920 & 0.810 & 0.665 & 0.093 & 0.511 & 0.893 \\
Internet Penetration Rate 2022 & 0.830 & 0.980 & 0.764 & 0.051 & 0.647 & 0.898 \\
Sex Ratio 2020 & 0.946 & 0.996 & 1.055 & 0.039 & 0.997 & 1.131 \\
Total Fertility Rate 2020 & 1.300 & 0.800 & 1.275 & 0.277 & 0.740 & 2.119 \\
\hline
\multicolumn{3}{l}{\textbf{Percentage of comments}} & \multicolumn{4}{c}{} \\
\multicolumn{2}{l}{\textit{Theoretical Dimensions}} & & & & & \\
Utility & 34.48\% & 51.18\% & 11.53\% & . & . & . \\
Childrearing Cost & 27.03\% & 16.03\% & 13.81\% & . & . & . \\
Values & 18.65\% & 13.19\% & 57.40\% & . & . & . \\
Pop. Structure & 14.73\% & 13.69\% & 15.00\% & . & . & . \\
Health & 5.11\% & 5.90\% & 2.25\% & . & . & . \\
\textit{Account Types} & & & & & & \\
 Personal Sharing & 57.57\% & 35.85\% & 48.28\% & . & . & . \\
Main Media & 19.98\% & 17.31\% & 5.29\% & . & . & . \\
Experts & 12.84\% & 4.25\% & 15.94\% & . & . & . \\
Bystanders & 9.61\% & 42.60\% & 30.49\% & . & . & . \\
\textit{Video stances} & & & & & & \\
Anti-reproduction & 57.58\% & 43.48\% & 69.58\% & . & . & . \\
Pro-reproduction & 42.42\% & 56.52\% & 30.42\% & . & . & . \\
\textit{Video figure gender} & & & & & & \\
Female & 61.12\% & 58.81\% & 54.66\% & . & . & . \\
Male & 38.88\% & 41.19\% & 45.34\% & . & . & . \\
\textit{Video figure age} & & & & & & \\
Youth & 41.89\% & 87.20\% & 31.56\% & . & . & . \\
Middle-aged & 51.09\% & 11.41\% & 48.59\% & . & . & . \\
Old & 7.02\% & 1.39\% & 19.85\% & . & . & . \\
\textit{Comment year} & & & & & & \\
2021 & 8.02\% & 6.39\% & . & . & . & . \\
2022 & 20.50\% & 14.61\% & 12.71\% & . & . & . \\
2023 & 38.74\% & 27.14\% & 32.28\% & . & . & . \\
2024 & 32.07\% & 51.87\% & 55.02\% & . & . & . \\
\hline
Obs. & 82,670 & 61,171 & 72,595 & & & \\
\hline
\multicolumn{7}{l}{Note: Because Japan and South Korea do not have sub-regions, } \\
\multicolumn{7}{l}{they do not have standard deviations of the regional context variables.} \\
\end{tabular}
}
\label{appendix:desc}
\end{table}


\end{document}